\documentclass[journal=jacsat,manuscript=article]{achemso}
\usepackage{multirow}
\usepackage{color,amsmath,amssymb,comment}
\setkeys{acs}{articletitle = true}

\newcommand{\figref}[1]{Figure~\ref{#1}}

\newcommand{\new}[1]{#1}
\newcommand{\rem}[1]{}

\usepackage[normalem]{ulem}

\title{Formation of Acetaldehyde on CO-rich Ices}

\author{Thanja Lamberts}
\email{a.l.m.lamberts@lic.leidenuniv.nl}
\affiliation{Institute for Theoretical Chemistry, University of Stuttgart, 
                Pfaffenwaldring 55, 70569 Stuttgart, Germany}
\altaffiliation{Leiden Institute of Chemistry, Gorlaeus Laboratories,
Leiden University, 
P.O. Box 9502, 
2300 RA Leiden, The Netherlands}

\author{Max N. Markmeyer}
\affiliation{Institute for Theoretical Chemistry, University of Stuttgart, 
                Pfaffenwaldring 55, 70569 Stuttgart, Germany}

\author{Florian J. Kolb}
\affiliation{Institute for Theoretical Chemistry, University of Stuttgart, 
                Pfaffenwaldring 55, 70569 Stuttgart, Germany}

\author{Johannes K\"{a}stner}
\email{kaestner@theochem.uni-stuttgart.de}
\affiliation{Institute for Theoretical Chemistry, University of Stuttgart, 
                Pfaffenwaldring 55, 70569 Stuttgart, Germany}

\begin{document}

\begin{abstract}
  The radicals HCO and CH$_3$ on carbon monoxide ice surfaces were simulated
  using density functional theory. Their binding energy on amorphous CO ice
  shows broad distributions, with approximative average values of 500~K for
  HCO and 200~K for CH$_3$. If they are located on the surface close to each
  other (3 to 4~\AA), molecular dynamics calculations based on density
  functional theory show that they can form acetaldehyde (CH$_3$CHO) or CH$_4$
  + CO in barrier-less reactions, depending on the initial orientation of the
  molecules with respect to each other. In some orientations, no spontaneous
  reactions were found, the products remained bound to the surface. Sufficient
  configurational sampling, inclusion of the vibrational zero point energy,
  and a thorough benchmark of the applied electronic structure method are
  important to predict reliable binding energies for such weakly interacting
  systems. From these results it is clear that complex organic molecules, like
  acetaldehyde, can be formed by recombination reactions of radicals on CO
  surfaces.
\end{abstract}


\section{Introduction} \label{sec:intro}

Acetaldehyde is a major precursor for several complex organic molecules (COMs)
in astrochemical surface reactions. It has first been detected in
  the interstellar medium in the 70's of
the previous century \cite{Gottlieb:1973, Fourikis:1974,Gilmore:1976} and
thereafter it has also been detected in other cold clouds, TMC-1 and L134N,
\cite{Matthews:1985} translucent clouds\cite{Turner:1999,
  Thiel:2017,Belloche:2013} and pre-stellar cores.\cite{Bacmann:2012,
  Vastel:2014} In CO rich ices on the surfaces of dust grains, the direct
hydrogenation of CO leads to the closed-shell molecules formaldehyde and
methanol. En route to these, radicals such as HCO are formed. Larger molecules
that exhibit an additional C--C bond are believed to be formed, at least
partially, by radical-radical recombination reactions at low
temperatures.\cite{Garrod:2006,Woods:2013}. In other words, when two radicals
find each other it is often assumed that a barrier-less reaction can take
place leading to the formation of larger closed-shell molecules. Radicals can
encounter each other either via creation in the ice next to each other,
\cite{Fedoseev:2015} via thermally induced diffusion\cite{Garrod:2006} or via
hot diffusion following reaction.\cite{Lamberts:2014b} The radical-radical
recombination of HCO and CH$_3$ can lead to acetaldehyde, but can also lead to
CH$_4$ + CO, i.e., no formation of a new C--C bond. For the similar reaction
HCO + NH$_2$ on a water ice surface it was recently shown that the outcome
depends on the orientation of the radicals with respect to each
other.\cite{rim18}

The solid-state formation of acetaldehyde has experimentally been confirmed
following UV-irradiation of methanol ices, \cite{Oberg:2009,
  Paardekooper:2016} albeit in low quantities. One possible reaction mechanism
is through the C--C bond formation via a reaction between CH$_3$ and HCO
radicals. This reaction, however, can also take place without additional
energetic input of UV-irradiation, as long as both radicals are
available. This may indeed be the case during the later stages of the cold
dense clouds, when more CO is available and hydrogenation of carbon monoxide
results in HCO formation. In either case, methanol photodissociation or
non-energetic processing, such a reaction is expected to take place in an ice
matrix that is no longer water dominated.\cite{Boogert:2015}

In this work we show that in a CO-rich environment indeed the formation of
acetaldehyde from HCO and CH$_3$ is possible. However, when these radicals
meet, another product channel leading to CH$_4$ + CO is also
open. Furthermore, we show that the outcome depends on the relative
orientation of the two radicals with respect to each other. This is the first
ab initio computational study of a radical-radical reaction on a CO ice
cluster. As such this is also a comment on the importance of taking into
account other ice constituents than only H$_2$O.

\section{Methodology} \label{sec:meth}

The methods to simulate the radical-radical reaction are comprised of three
steps: 1) the construction of a CO cluster, 2) the determination of HCO and
CH$_3$ binding sites and corresponding binding energies, and 3) the
direct simulation of the radical-radical surface
reactions HCO + CH$_3$. For all calculations density functional theory
(DFT) has been used within the program
ChemShell.\cite{Sherwood:2003,Metz:2014} Both molecular dynamics (MD) runs and
geometry optimizations were carried out within this framework. Geometry
optimizations were performed with the DL-find library.\cite{Kaestner:2009} The
PBEh-3c functional in combination with the def2-mSVP basis
set\cite{Grimme:2015} in turbomole\cite{TURBOMOLE} are used for all MD runs as
well as for the geometry optimizations. Refinements of the binding sites and
binding energies are also calculated with the M06-2X
functional\cite{Zhao:2008} and def2-TZVPD basis
set\cite{Weigend:1998,Rappoport:2010} including a Grimme's D3
correction.\cite{Grimme:2010} Finally we provide a small benchmark study at
the CCSD(T)-F12/VTZ-F12 level of theory\cite{Knowles:1993,
  Deegan:1994, Knizia:2009,Adler:2007, Peterson:2008} using molpro\cite{werner2012} to get a feeling for the
error bar on the binding energies.

\begin{figure}[htbp!]
\centering
\includegraphics[width=8cm]{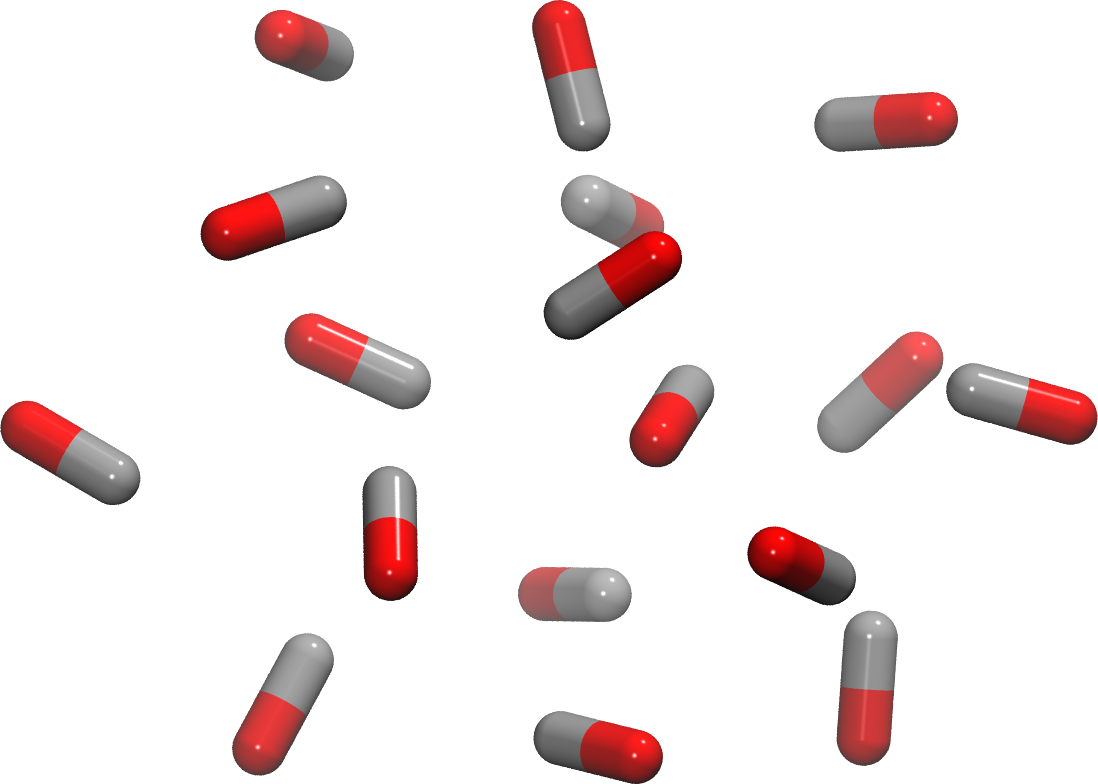} 
\caption{The bare CO cluster used in this study.\label{fig:cluster}}
\end{figure}

\subsection{Construction of the CO cluster}
A cluster consisting of 17 CO molecules was prepared using
Avogadro\cite{Avogadro} and subsequently relaxed via a short ab initio MD run
at the PBEh-3c/mSVP level of theory, followed by geometry optimization. In
this way, an amorphous ellipsoidal structure of approximately
$12\times 8 \times 8$ {\AA}$^3$ is obtained, allowing for a variety of binding sites that
is used throughout the calculations. The geometry of the resulting cluster is
shown in \figref{fig:cluster}.

\subsection{HCO and CH$_3$ binding sites}
For each radical, HCO and CH$_3$, the binding site distribution was probed by
generating 60 different positions of the radical arranged symmetrically
  at the vertices of a truncated icosahedron at a distance of
approximately 3--4 {\AA} above the surface around the entire cluster. Geometry
optimization at the PBEh-3c/mSVP level of theory was then performed resulting
in 60 binding sites. Binding energies were determined via the difference in
energy between the adsorbed species on the cluster and the separated radical
and cluster.  In other words, the relaxation of the surface is included in the
binding energy. We used the convention that positive binding
  energies refer to exothermic binding. Subsequently, binding energies at the M06-2X/TZVPD+D3 level of
theory were calculated of the previously optimized structures.

Various different binding modes for both radicals were found. Three
representative geometries were selected for HCO and two for CH$_3$ for which
the harmonic zero-point vibrational energy (ZPE) correction to the binding
energy was calculated with PBEh-3c/mSVP. The coordinates of these
representative structures are provided in the Supporting Information. These
specific five geometries were then further refined through a reoptimization
with M06-2X/TZVPD+D3. During this reoptimization, to reduce the computational
cost, all but the four CO molecules closest to the adsorbant were kept frozen.

\subsection{Surface reaction of HCO + CH$_3$}

The surface reaction of HCO with CH$_3$ was studied through MD simulations in
the microcanonical ensemble (NVE). Low random initial velocities,
corresponding to a temperature of 10~K, were assigned to restrict the study to
barrier-less processes. A small time step of 0.2~fs was used. Overall, 56
trajectories were run, for 150 steps each (another 150 steps if no reaction
was found within the first 150 steps). The electronic structure of the
biradical at the PBEh-3c/mSVP level was an unrestricted broken symmetry wave
function that showed clear separation of the spin densities between the
radicals, positive spin density on one radical and negative spin density on
the other one.  The five previously selected adsorbed radical geometries were
chosen as starting points and for each of these the other radical was placed
as relaxed structure in close proximity, 3--5 {\AA} in various starting
geometries. Because the interaction between the radicals and the surface
  is rather weak, we assume that these represent structures accessible by
  diffusion. In very few cases, non-physical structures were obtained, in
  which the reactant turned out to be too close to the CO cluster
  initially. In these cases it flew away rather than reacting. These were not
  considered in the following. Here, we show that for each binding site both
product channels are available, \emph{i.e.}, CH$_3$CHO and CH$_4$ + CO can be
formed. Additionally an outcome of the simulation can also be that no reaction
takes place.

In principle it is possible to obtain a branching ratio for the two different
product channels by sampling sufficient binding sites per binding mode of the
first adsorbed radical and running MD simulations for multiple starting
geometries with the second radical in close proximity. However, this falls
beyond the scope of this work as too many simulations would be necessary to
yield a statistically relevant result.

\subsection{Benchmark of the level of theory}

Given the total system size, between 34 and 41 atoms, we are restricted to
using mainly the PBEh-3c/mSVP level of theory. Often the M06-2X functional is
seen as a more accurate choice when higher levels of theory, such as
CCSD(T)-F12, are not available.\cite{Sameera:2017,Wakelam:2017} In
Table~\ref{benchmark} the interaction energies of several smaller models,
CH$_3$--CO and HCO--CO dimers and CH$_3$--(CO)$_2$ and HCO--(CO)$_2$ trimers,
are presented to estimate the accuracy for the larger cluster used in the
  rest of this work. The values for PBEh-3c and M06-2X calculations are
obtained following a full optimization routine, whilst the values at the
CCSD(T)-F12 level of theory refer to energies calculated at the geometries
optimized with PBEh-3c and M06-2X. All values are given without zero-point
energy correction.

\begin{table*}[t]
\centering
\caption{\textbf{Benchmark of interaction energies (in K) of CO with CH$_3$
    and HCO.}}
\label{benchmark}
\begin{tabular}{l|rr|rr}
						& 	PBEh-3c 	&     CCSD(T) & M06-2X & CCSD(T) \\ 
						& 			& /PBEh-3		& 		& /M06-2X \\
\hline
HCO-CO a				& 	320		& 	250		& 	245			& 250	\\
HCO-CO b				& 	890		& 	485		& 	245			& 250	\\
HCO-(CO)$_2$ a	 		& 	1100	& 	445		& 	750			& 550	\\
HCO-(CO)$_2$ b 			& 	1400	& 	630		& 	820			& 740	\\
CH$_3$-CO 				& 	495		& 	110		& 	665			& 125	\\
CH$_3$-(CO)$_2$ a	 	& 	1355	& 	440		& 	1045		& 515	\\
CH$_3$-(CO)$_2$ b 		& 	1000	& 	225		& 	1313		& 365	\\
\end{tabular}\\
The labels a and b refer to different conformers.
\end{table*}

It is immediately clear from the comparison between the DFT and CCSD(T)-F12
values that both functionals tend to overestimate the interaction energy,
predict too strong binding. This increases with the number of CO molecules in
the cluster and is largest for the CH$_3$-(CO)$_{n}$ systems. In other words,
\new{the binding energy distributions obtained should be shifted somewhat to
  lower binding energies.}  
\rem{assuming that the radicals on the surface
  interact with at least two CO molecules, the binding energy distributions
  obtained should be shifted by roughly 700~K for those at the PBEh-3c level
  of theory and by roughly 200 (HCO) or 600~K (CH$_3$) for those at the M06-2X
  level of theory.}

%%%%%%%%%%%%%%%%%%%%%%%%%%%%%%%%%%%%%%%%%%%%%%%%%%%%%%%%%%%%%%%%%%%%%%%%%%%%%
\section{Results and Discussion} \label{sec:rd}
%%%%%%%%%%%%%%%%%%%%%%%%%%%%%%%%%%%%%%%%%%%%%%%%%%%%%%%%%%%%%%%%%%%%%%%%%%%%%

\subsection{HCO and CH$_3$ binding sites}

\begin{table*}[htbp!]
\centering
\caption{\textbf{Ranges of binding energies of HCO and CH$_3$ to amorphous CO
    (this work) compared to water ice.}}\label{comparisonsameera}
\begin{tabular}{l|cc|cccc}
		& \multicolumn{2}{c|}{amorphous CO} & \multicolumn{2}{c}{crystalline H$_2$O$^*$} & single H$_2$O$^\dag$ & amorphous H$_2$O$^\ddag$\\
		& M06-2X& PBEh-3c 		& M06-2X & wB97XD & M06 \& MP2 & M06-2X+D3\\
\hline
HCO 	& 230--2300 	& 500--2100	& 1400--4900 & 350--5000  & 2400 & 2333 \\
avg.	& 1050		& 1250		& & \\
CH$_3$	& 175--1900 	& 250--1300 	& 1300--3100 & 700--3250  & 1600 & 734 \\
avg.	& 1150		& 900		& & \\
\multicolumn{7}{l}{$^*$ \citet{Sameera:2017}, $^\dag$ \citet{Wakelam:2017}, $^\ddag$ \citet{Enrique-Romero:2016}}
\end{tabular}
\end{table*}

Binding energies for 60 binding sites for HCO and CH$_3$ on the CO cluster
were calculated. The average potential energy for binding was $1050$~K for
HCO and $1150$~K for CH$_3$, both obtained with M06-2X+D3, see
Table~\ref{comparisonsameera}. With PBEh-3c/mSVP the binding is predicted
somewhat stronger for HCO and somewhat weaker for CH$_3$, see also
Table~\ref{comparisonsameera}.  Depending on the binding sites, the binding
energies are rather broadly distributed, as indicated in
Table~\ref{comparisonsameera}. This is consistent with findings for other
adsorbates on amorphous surfaces.\cite{son16,lam17a,son17} Histograms of the
binding energy distributions are provided in the Supporting Information.

These values are compared to the binding energies to water calculated by
\citet{Sameera:2017} for the same radicals on a crystalline water surface, to
the results by \citet{Wakelam:2017} obtained from interaction with a single
water molecule, and to the results by \citet{Enrique-Romero:2016} on an
amorphous water cluster in Table~\ref{comparisonsameera}. The interaction of
both radicals with a H$_2$O surface is clearly stronger than with a CO
surface, as expected.

\begin{figure}[htbp!]
\centering
\includegraphics[width=8cm]{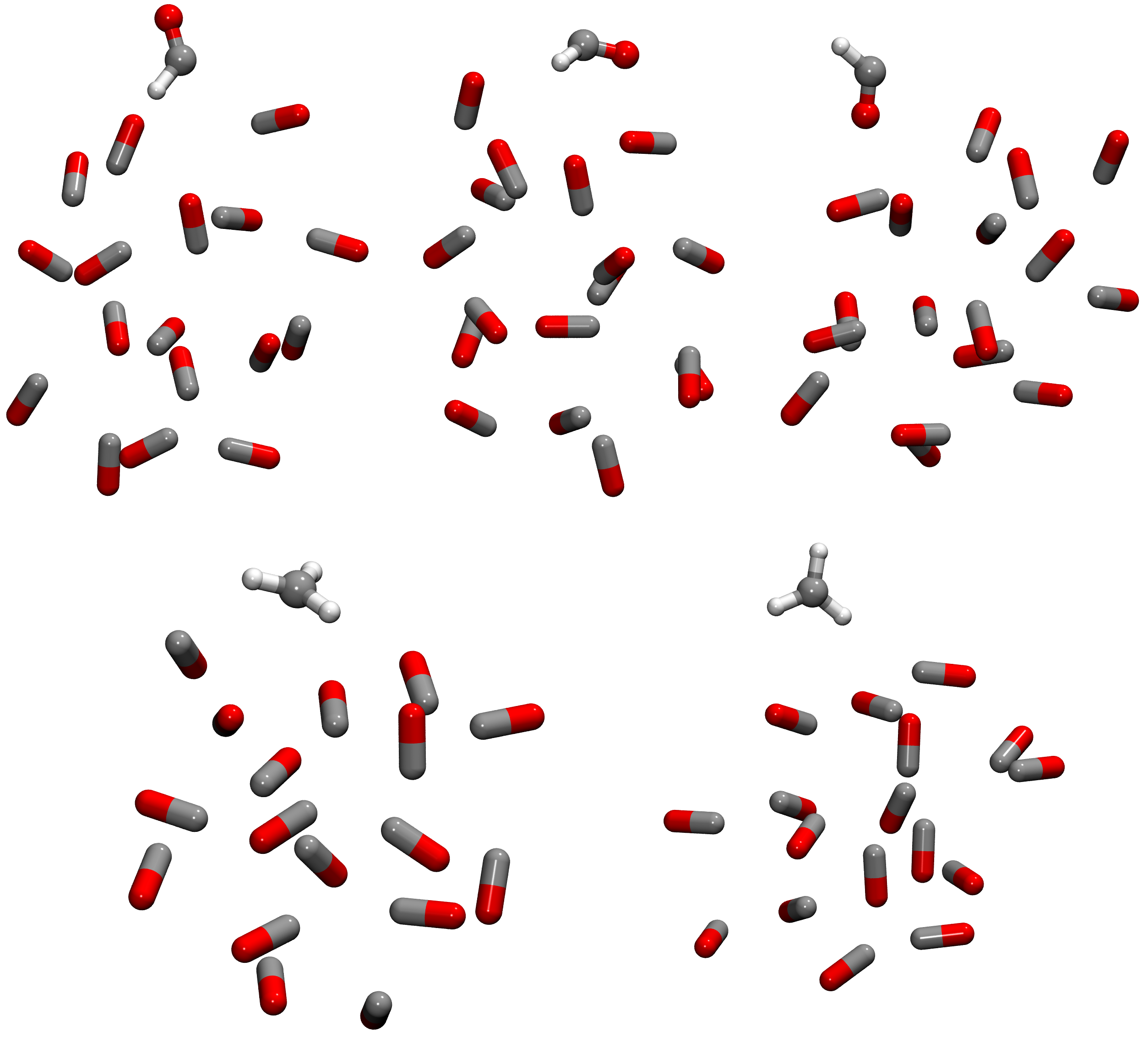}
\caption{Binding modes of HCO (top) and CH$_3$ (bottom) to the CO cluster. The
  order corresponds to Table~\ref{comparisoninclzpe}.\label{bindingmodes}}
\end{figure}

\begin{table}[htbp!]
\centering\small
\caption{\textbf{Binding energies (in K) of HCO and CH$_3$ on specific sites
    of a CO cluster.}}\label{comparisoninclzpe}
\begin{tabular}{lr@{}lrrr}
				& M06-2X+D3&	 & \multicolumn{3}{c}{PBEh-3c}  \\
				& $E_\text{bind}$&  & $E_\text{bind}$ & ZPE	& Total	 \\
\hline
HCO-H 			& 915&			& 1070	& $-$350	& 720 	 \\
HCO-flat 			& 1180&			& 1410	& $-$480	& 930	 \\
HCO-O			& 1480&$^{a}$	& 1250	& $-$270	& 980	 \\
CH$_3$-flat		& 1960&			& 1140	& $-$450	& 690	 \\
CH$_3$-H 		& 845&$^{a}$		& 630	& $-$270	& 360	 \\
\end{tabular}\\
$^{a}$ The reoptimization with M06-2X+D3 lead to a considerable change in the geometry
\end{table} 

\begin{table}[htbp!]
\centering\small
\caption{\textbf{Non-bonded interatomic distances (in \AA) of typical binding sites of HCO and
    CH$_3$ on a CO cluster.}}\label{dinstances}
\begin{tabular}{lrrrrrrr}
   & H--O & H--O & H--O & O--C & O--C \\
\hline
HCO-H 			& 2.91 & 3.05 & 3.18 \\
HCO-flat 		& 3.03 & 3.04 & 3.07 \\
HCO-O			& & & & 3.17 & 3.18 \\
\hline
 &  C--C & C--C & H--C & H--O\\
\hline
CH$_3$-flat		& 3.42 & 3.53 \\
CH$_3$-H 		& & & 2.78 & 3.14 \\
\end{tabular}\\
\end{table}

Five representative binding modes were analyzed in more detail. The ZPE was
calculated on the PBEh-3c/mSVP level. Subsequently, the geometry was refined
at the M06-2X/TZVPD+D3 level. The resulting binding energies are listed in
Table~\ref{comparisoninclzpe}.  The ZPE correction is on average around 360~K
for both radicals, but varies a bit between the binding sites. The structures
are shown in \figref{bindingmodes}, and non-bonded interatomic distances
between the adsorbates and the CO cluster are given in Table~\ref{dinstances}.
\new{All of the values are distances between the adsorbate and the nearest CO
  molecules of the cluster. The label O--C refers to distances from O of HCO
  to C of the nearest CO molecules, C--C refers to the distance of C of CH$_3$
  to C of CO. It is clear from the rather large distances, that the adsorbates
  interact rather weakly with the cluster. The full coordinates of these
  representative structures are provided in the Supporting Information.}

To estimate total binding energies for HCO and CH$_3$, the average binding
energy on the M06-2X+D3 level, the ZPE correction and the shift between M06-2X and
CCSD(T)-F12 should be taken into account. These result in very rough estimates
of average binding energies of $500$~K for HCO and $200$~K for CH$_3$ on a
CO surface. It should be noted that we find a significant variation between
our levels of theory, but even stronger variations between the different
binding sites. However, it can be concluded that the interaction of CO with
adsorbed HCO or CH$_3$ is so weak that diffusion might actually be easy on
CO-rich surfaces as our binding energies are in the same range as the
diffusion barriers obtained
experimentally for H on CO \cite{Kimura:2018}.

\subsection{Surface reaction of HCO + CH$_3$}

For each of the five representative geometries %mentioned in Table~\ref{comparisoninclzpe} 
depicted in \figref{bindingmodes} 
between 4 and 6 different placements of the second radical were used as
initial geometries for MD simulation runs. For each of these sets, we observed
both reactive channels, to CH$_4$ + CO as well as to CH$_3$CHO. These are
indeed barrier-less. The channel observed depends on the orientation of the
two radicals with respect to each other.
In some cases, no reaction was found. The chosen starting geometries probably
lead to small barriers in these cases.

\begin{figure}[htbp!]
\centering
\includegraphics[width=8cm]{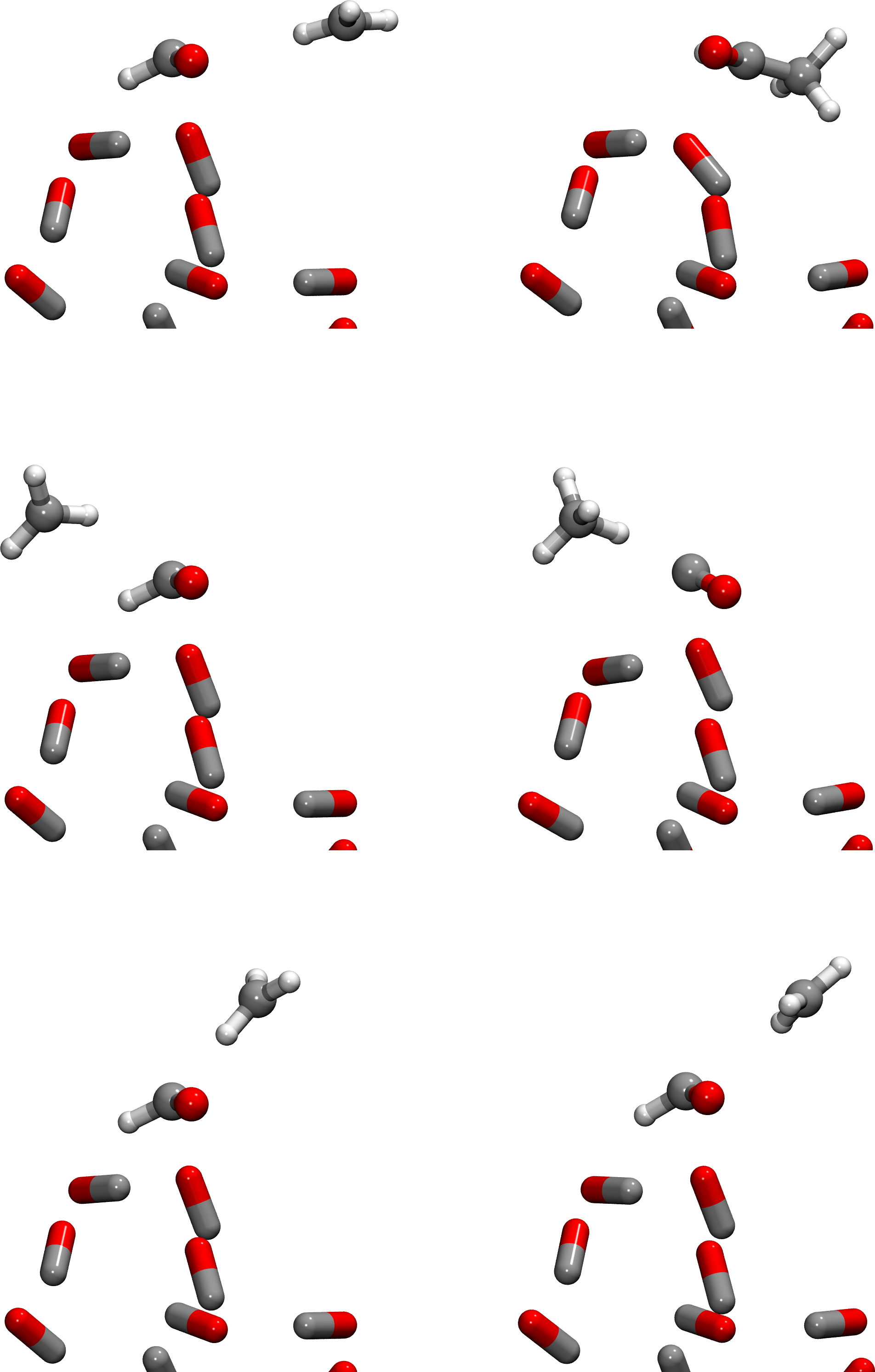} 
\caption{Representative examples of pathways of HCO and CH$_3$ leading to
  CH$_3$CHO (top), CH$_4$ + CO (middle) or non-reactive binding of both
  reactants to the CO cluster (bottom).\label{HCO-flat}}
\end{figure}

\figref{HCO-flat} shows the initial geometries (left) and a representative
snapshot after the reaction (right) of the MD runs. The CO molecules are shown
as sticks, the reactive radicals and their products as balls and sticks. These
three trajectories were obtained from the same initial binding mode of HCO
bound flat on the surface. The CH$_3$ moiety was placed at different positions,
leading to different products. Note that given the low binding energies of the
radicals with the cluster, the initial geometries, for which the second
reactant seems to be somewhat distant from the cluster, are actually
realistic. Here we do not consider the reaction of one adsorbed radical with
another radical from the gas phase.

For each of the five selected geometries
(\figref{bindingmodes}) both CH$_4$ + CO and CH$_3$CHO can be formed.  In
several cases, the reactants stayed associated with the CO cluster, but did
not react.

\section{Conclusions} \label{sec:conc}

We have studied the chemistry of HCO and CH$_3$ radicals on an amorphous CO
ice surface. The binding energies of both radicals are low, which is
understandable given their low dipole moments and the inert character of the CO
surface. The binding energies of  approximately 500~K for
  HCO and 200~K for CH$_3$ are comparable to those of hydrogen atoms on the
CO surface. As found for other adsorption processes on amorphous surfaces, a
broad distribution of the binding energies of HCO and CH$_3$ was obtained. The
comparison of the levels of theory demonstrated that a thorough benchmarking,
the inclusion of vibrational zero point energy, as well as sufficient
configurational sampling are necessary for reliable prediction of binding
energies. When both radicals are placed close to each other (3 to 4~{\AA}),
the relative orientation of the reactants with respect to each other
determines which of the products is formed: (a) CH$_4$ + CO, (b) CH$_3$CHO, or
(c) no reaction. Thus, we have clearly shown that the formation of complex
organic molecules, like acetaldehyde, is possible from simpler radical species
like HCO and CH$_3$ on a CO surface. We also found that some reaction products
can desorb from the surface. This will, however, depend strongly on the
morphology of the ice. On a rougher surface, the molecules are more likely to
thermalize and stay bound to the ice. Overall, this study should serve as a
proof of principle that radical-radical recombinations can lead to the buildup
of complex organic molecules rather than as a quantitative study, which would
require much more configurational sampling. We hope that this will inspire
other theoretical chemists to investigate further similar combinations, which
may explain the buildup of larger, more complex organic species in interstellar environments.

\section*{Acknowledgments}
 Albert Rimola is thanked for fruitful and stimulating discussions.
The  authors  acknowledge  support  for  computer  time  by the state of Baden-W\"{u}rttemberg through bwHPC and the
Germany  Research  Foundation  (DFG)  through  grant  no. INST 40/467-1FUGG. This project was financially supported by the European Union’s Horizon 2020 research  and  innovation  programme  (grant  agreement  No.
646717,  TUNNELCHEM),  the  Alexander  von  Humboldt Foundation, the Netherlands Organisation for Scientific Research (NWO) via a VENI fellowship (722.017.008) and the COST Action CM1401 via an STSM travel grant.

\section*{Supporting Information}

Cartesian coordinates of the CO cluster, of representative binding modes, and histograms of
the binding energy distributions are provided free of charge on the ACS
Publications website.

\bibliography{biblio}

\newpage 
\section*{Table of Contents only:}
\includegraphics[width=8.25cm]{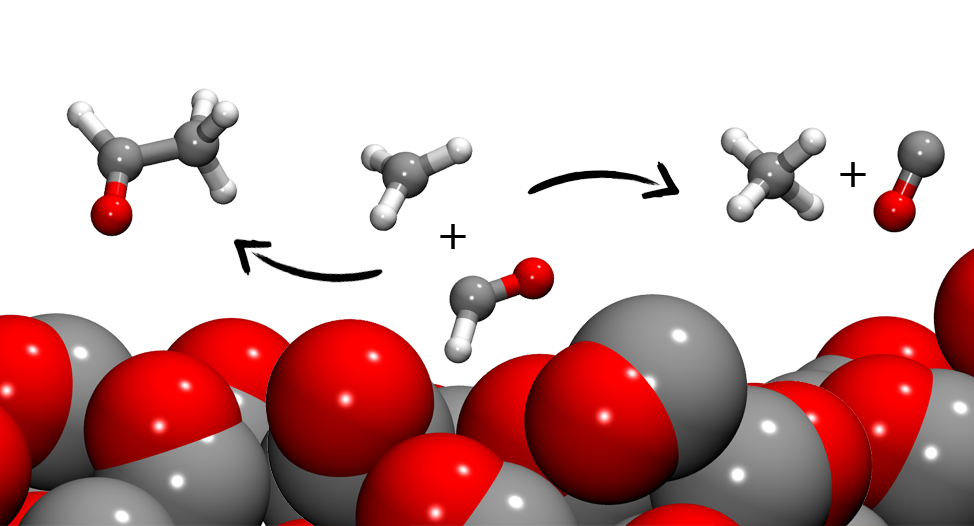}

\end{document}